\documentclass[]{aastex}
\usepackage{txfonts,amssymb,graphicx,color}
\begin{document}
\title{Evidence from the asteroid belt for a violent past evolution of Jupiter's orbit}

\author{\textbf{Alessandro Morbidelli}} \affil{\small\em Departement Cassiop\'{e}e: Universite de
Nice - Sophia Antipolis, Observatoire de la C\^{o}te d'Azur, CNRS 4, 06304 Nice, France}

\author{\textbf{Ramon Brasser}} \affil{\small\em Departement Cassiop\'{e}e: Universite de
Nice - Sophia Antipolis, Observatoire de la C\^{o}te d'Azur, CNRS 4, 06304 Nice, France}

\author{\textbf{Rodney Gomes}} \affil{\small\em Observatório Nacional, Rua General José Cristino 77, CEP 20921-400, Rio de Janeiro, RJ,
Brazil}  

\author{\textbf{Harold F$.$ Levison}} \affil{\small\em Southwest Research Institute, 1050 Walnut St, Suite 300, Boulder, CO 80302, USA}
\and

\author{\textbf{Kleomenis Tsiganis}} \affil{\small\em Department of Physics, Aristotle University of Thessaloniki, 54 124 Thessaloniki, Greece}

\received{by AJ, on may 24, 2010}
\accepted{by AJ, on Sept 8, 2010}

\begin{abstract}
We use the current orbital structure of large ($>50$~km) asteroids in
the main asteroid belt to constrain the evolution of the giant planets
when they migrated from their primordial orbits to their current
ones. Minton \& Malhotra (2009) showed that the orbital distribution
of large asteroids in the main belt can be reproduced by an
exponentially-decaying migration of the giant planets on a time scale
of $\tau \sim 0.5$~My. However, self-consistent numerical simulations
show that the planetesimal-driven migration of the giant planets is
inconsistent with an exponential change in their semi major axes on
such a short time scale (Hahn \& Malhotra, 1999). In fact, the typical
time scale is $\tau \ge 5$~My.  When giant planet migration on this
time scale is applied to the asteroid belt, the resulting orbital
distribution is incompatible with the observed one. However, the
planet migration can be significantly sped up by planet-planet
encounters. Consider an evolution where both Jupiter and Saturn have
close encounters with a Neptune-mass planet (presumably Uranus or
Neptune themselves) and where this third planet, after being scattered
inwards by Saturn, is scattered outwards by Jupiter. This scenario
leads to a very rapid increase in the orbital separation between
Jupiter and Saturn that we show here to have only mild effects on the
structure of asteroid belt. This type of evolution is called a
jumping-Jupiter case. Our results suggest that the total mass and
dynamical excitation of the asteroid belt before migration were
comparable to those currently observed. Moreover, they imply that,
before migration, the orbits of Jupiter and Saturn were much less
eccentric than the current ones.
\end{abstract}


\section{Introduction}
This paper is the third in a series in which we try to unveil the past
evolution of the orbits of the giant planets using the current
dynamical structure of the Solar System. In these works, we do not
assume a priori any preferred model other than the well accepted fact
that the giant planets were much closer to each other in the past and
somehow moved towards their current orbital radii.  This radial
migration of the giant planets is believed to be the last major event
that sculpted the structure of the Solar System (Levison et al., 2001;
Gomes et al., 2005; Strom et al., 2005), an assumption which is
implicit in our works.\\

In two previous studies, Morbidelli et al. (2009) and Brasser et
al. (2009), henceforth labelled Paper I and Paper II respectively, we
investigated how the giant planets (Paper I) and the terrestrial
planets (Paper II) could have achieved orbits with their current
secular properties i.e. with their current frequencies and amplitudes
in eccentricity and inclination. The orbits of the planets are not
fixed ellipses: they undergo precession, and the eccentricities and
inclinations have long-term (secular) oscillations that, to a first
approximation, are described by the Lagrange-Laplace theory:
\begin{eqnarray}
e_k\exp({{\iota}}\varpi_k)=\sum_j M_{j,k}\exp[{{\iota}}(g_j t + \beta_j)],\nonumber \\
\sin(i_k/2)\exp({{\iota}}\Omega_k)=\sum_j N_{j,k}\exp[{{\iota}}(s_j t + \beta_j)].
\label{Lagrange}
\end{eqnarray}
Here $\iota$ is the immaginary unit and $e_k, i_k, \varpi_k, \Omega_k$
are, respectively the eccentricity, inclination, longitude of
perihelion and longitude of the node of the planet $k$, $g_j$ and
$s_j$ are the secular frequencies of the system (the index $j$ running
over the number of planets) and the coefficients $M$ and $N$ are the
secular amplitudes. \\

In Paper I we concluded that close encounters between the giant
planets is the only known mechanism that can explain the current
amplitude of $M_{5,5}$ i.e. the excitation of Jupiter's eccentricity
associated with the $g_5$ secular frequency. Other possible
mechanisms, such as the crossing of multiple mean-motion resonances
between Jupiter and Saturn, only excite $M_{5,6}$. Resonance crossings
between Saturn and Uranus (which also kick Jupiter's orbit) are not
strong enough to pump $M_{5,5}$ to its current value. However, a
Neptune-mass planet (presumably Neptune itself or Uranus) encountering
Saturn in general excites $M_{5,5}$ to values comparable to the
current one.  In principle, encounters of a planet with Jupiter do not
need to have occurred. Planet-planet encounters are consistent with
the giant planet evolution model of Thommes et al. (1999) and with the
so-called ``Nice model'' (Tsiganis et al, 2005; Gomes et al.,
2005). \\

In Paper II we focused our attention on the terrestrial planets. We
showed that if the terrestrial planets were initially on
quasi-circular, nearly-coplanar orbits, the divergent migration of
Jupiter and Saturn must have been fast, otherwise the orbits of the
terrestrial planets would have become too eccentric and/or
inclined. Their excitation during giant planets migration is primarily
caused by the crossing of the secular resonances $g_5=g_2$ and
$g_5=g_1$, which pump $M_{2,k}$ and $M_{1,k}$. These resonances occur
because $g_5$ decreases while the orbital separation between Jupiter and Saturn
increases. More precisely, assuming that the orbital separation
between Jupiter and Saturn increased smoothly as in Malhotra (1993,
1995) by
\begin{equation}
\Delta a(t)=\Delta a_{\rm now}-\Delta_0\exp(-t/\tau),
\label{renu}
\end{equation} 
then $\tau$ had to be shorter than $\sim$1~My in order for the
terrestrial planets not to become too eccentric.  We were concerned by
such a short timescale, for the reasons detailed in sect.~\ref{migr}.
So, in Paper II we identified a plausible scenario that resulted in a
divergent radial displacement of the orbits of Jupiter and Saturn that
is rapid enough, although not of the smooth exponential form given in
eq.~(\ref{renu}), that we called the 'jumping-Jupiter' scenario. Here
a Neptune-mass planet is first scattered inwards by Saturn and then
outwards by Jupiter, so that the two major planets recoil in opposite
directions. However, we were unable to firmly conclude that the real
evolution of the giant planets had to be of the jumping-Jupiter type
because of the following hypothetical alternative. Suppose that after
their formation the terrestrial planets had more dynamically excited
orbits than now (particularly with larger amplitudes $M_{1,k}$ and
$M_{2,k}$). Their eccentricities could have been damped by the same
mechanism that would have excited them if they had been nearly
zero. In other words, the passage through the secular resonances
$g_5=g_2$ and $g_5=g_1$, could have decreased the values of $M_{2,k}$
and $M_{1,k}$ to the current values, provided these resonance passages
had occurred with the appropriate phasing.\\

The discussion on whether the orbital separation between Jupiter and
Saturn increased smoothly, as in eq.~(\ref{renu}) or abruptly (as in
the jumping-Jupiter scenario) is not only of academic interest. These
two modes of orbital separation correspond to two radically different
views of the early evolution of our Solar System. In the first case,
such evolution was relatively smooth, and the increase in orbital
separation of the gas giants was driven solely by planetesimal
scattering. In the jumping-Jupiter scenario, Jupiter was involved in
encounters with another planet. In this case, the evolution of the outer solar
system would have been very violent, similar to the one that is
expected to have occurred in many (or most) extra-solar planetary
systems. \\

In this work, we turn our attention to the asteroid belt. Similar to
the terrestrial planet region, the migration of the giant planets
drives secular resonances through the belt, but now these are $g=g_6$
and $s=s_6$ ($g$ and $s$ denoting generically the pericenter and nodal
precession frequencies of the asteroids, while $g_6$ and $s_6$ are the
mean precession frequencies of Saturn). The radial displacement of
these secular resonances affects the asteroids' local orbital
distribution in a way that depends sensitively on the rate of
migration (Gomes, 1997). Therefore, reproducing the current orbital
distribution of the asteroid belt under different conditions can lead
to strong constraints on how Jupiter and Saturn separated from each
other. In addition the orbital properties of the asteroid belt might
also provide information about the orbital configuration of the giant
planets prior to their migration: it might help us constrain whether
the pre-migration orbits of the giant planets were more circular or
eccentric.\\

Recently, Minton \& Malhotra (2009) showed that the orbital
distribution of the asteroid belt is consistent with a smooth increase
in the separation between Jupiter and Saturn. They assumed that,
originally, the asteroids in the primordial belt were uniformly
distributed in orbital parameter space and that the separation between
the two gas giants increased as in eq.~(\ref{renu}), with
$\Delta_0=1.08$~AU (Malhotra, 1993) and $\tau=0.5$~My. Unfortunately,
different values of $\tau$ were not tested in that study nor did the
authors offer any suggestions for a possible mechanism for this fast
migration. Thus, in the first part of this paper we revisit Minton \&
Malhotra's work, with the aim of determining whether a value of $\tau$
as short as 0.5~My is really needed and what this implies. In
sect.~\ref{migr} we summarize the basic properties of planetesimal-driven
migration that are important for this problem. In
sect.~\ref{AB-smooth} we investigate the evolution of the asteroid
belt in case of a smooth, planetesimal-driven migration of Jupiter and
Saturn. After we demonstrate the incompatibility between the orbital
structure of the asteroid belt and this kind of migration, we describe
the jumping-Jupiter scenario in sect.~\ref{JJ}, followed by a
presentation of its effects on the orbital structure of the asteroid
belt in sect.~\ref{JJast}. The implications of this scenario are
discussed in sect.~\ref{imply}, and the conclusions follow in
sect.~\ref{concl}. {In Appendix, we summarize the sequence of the
  major events that characterized the evolution of the solar system,
  as emerging from this and other works.} 

\section{Brief description of planetesimal-driven migration}
\label{migr}

Planetesimal-driven migration occurs when a planet encounters a large
swarm of planetesimals. The planet scatters the planetesimals away
from its vicinity, which causes the planet and planetesimals to
exchange energy and angular momentum and thus the planet migrates (Fernandez
\& Ip, 1984; Malhotra, 1993, 1995; Ida et al., 2000; Gomes et al.,
2004; Kirsh et al., 2009). \\

For the planets in our solar system, numerical simulations
(e.g. Fernandez \& Ip, 1984; Hahn and Malhotra, 1999; Gomes et al.,
2004) show that Jupiter migrates inwards, while Saturn, Uranus and
Neptune migrate outwards.  The orbital separation between each pair of
planets increases with time. Over the dynamical lifetime of each
particle, each planet suffers a small change in its orbital semi major
axis, $\delta a$, which is different from planet to planet. Numerical
simulations of the Centaur population\footnote{The Centaurs are the
population of objects that are currently crossing the orbits of the
giant planets; this population can be considered as a proxy for the
primordial population of planetesimals that drove planet migration}
show that said population decays roughly exponentially with a certain
e-folding time, $\tau_C$ (e.g. DiSisto \& Brunini, 2007). Thus the
radial displacement of the planets must also decay exponentially, with
the same $\tau$:
\begin{equation}
a(t)=a_{\rm now}-\Delta a\,\exp(-t/\tau_C).
\label{migeq}
\end{equation}
Here $a(t)$ is the semi major axis of each planet as a function of
time, $a_{\rm now}$ the semi major axis of the planet at the end of
migration (i.e. the current semi major axis) and $\Delta a$ the total
radial distance that the planet migrates. \\

Strictly speaking, the identity between the planet migration timescale
$\tau$ and the planetesimals lifetime $\tau_C$ is valid only in case
of {\it damped migration} (Gomes et al., 2004).  In damped migration,
the loss of planetesimals is not compensated by the acquisition of new
planetesimals into the planet-scattering region that is due to the
displacement of the planet itself. Thus, planet migration slows down
progressively as planetesimals are depleted. In massive planetesimals
disks, planet migration can be {\it self-sustained} (Ida et al., 2000;
Gomes et al., 2004). In this case a planet can migrate through the
disk by scattering planetesimals and leaving them ``behind'' relative
to its migration direction. In this case, the migration speed does not
slow down with time: it actually accelerates (until some saturation or
migration reversal point is hit). So, obviously, a formula like
(\ref{migeq}) does not apply.  Gomes et al. (2004) made a convincing
case that self-sustained migration can not have occurred in the solar
system. Even if it had occurred, though, it would have concerned only
Neptune and Uranus. Jupiter and Saturn, given their large masses,
always have damped migration (unless one considers planetesimals disks
of unrealistic large masses, well exceeding the total mass of the gas
giant planets).  Thus,
for Jupiter and Saturn, we can assume with confidence that their
migration followed eq.(\ref{migeq}). Thus, their orbital separation
had to evolve as in eq.~(\ref{renu}), with $\tau=\tau_C$. \\

The parameter $\Delta a$ in (\ref{migeq}) (or $\Delta_0$ in
eq.~\ref{renu}) is related to the total amount of mass {of
the planetesimals that drive the planet migration. Thus, increasing
the mass of the planetesimal disk increases both the migration range
($\Delta a$) and the migration speed ($da/dt=\Delta a
[\exp(-t/\tau_C)]/\tau_C$).  However, when the planet is on an orbit
with a given semi major axis $\bar{a}$, its migration rate is
$da/dt(\bar{a})=(a_{\rm now}-\bar{a})/\tau_C$, i.e. it is {\it
independent} of $\Delta a$; therefore it is {\it independent} of the
mass of the planetesimal disk: it depends only on $\tau_C$. 

Obviously, the same is true for the speed of migration of a resonance
through a given location. A given location of a resonance corresponds
to a given semi major axis $\bar{a}$ of a planet. Thus, the speed at
which a resonance passes through the considered location, which in
turns determines its effects, depends
solely on the migration rate of the planet at $\bar{a}$ which, as we
have just seen, depends only on $\tau_C$ and not on the mass of the
planetesimal disk. We are lucky that this is the case because 
$\tau_C$ is pretty well constrained by simulations, whereas the total
mass of the planetesimal disk is open to speculation.\\ 

The value of $\tau_C$ for Centaurs found in the literature ranges from
6~My for objects the Jupiter-Saturn region (Bailey \& Malhotra, 2009) to
72~My in the Uranus-Neptune region (DiSisto \& Brunini, 2007). Thus,
we do not expect that planetesimal-driven migration of the giant
planets can occur with a value of $\tau$ significantly shorter than
$\sim$6~My. Indeed, a literature search for self-consistent
simulations of planetesimal-driven migration yields a typical time
scale $\tau\sim 10$~My (Hahn \& Malhotra, 1999; Gomes et al.,
2004). \\

By assuming that the orbital separation of Jupiter and Saturn evolved
as in eq.~(\ref{renu}), Minton \& Malhotra (2009) adopted a functional
form that is appropriate for planetesimal-driven migration. However,
the value of $\tau$ that they assumed (0.5~My), is not. The
relevant $\tau$ (i.e. $\tau_C$) is 10--20 times longer. For this
reason, below we repeat the calculation of Minton \& Malhotra (2009)
using $\tau \sim 5$~My.

\section{Planetesimal-driven migration and the asteroid belt}
\label{AB-smooth}

In this section we discuss the effects of planetesimal-driven
migration of the giant planets on the asteroid belt. Our goal is to
understand whether this kind of migration could have left the asteroid
belt with an orbital structure compatible with the current one, for a
reasonable set of initial conditions of the asteroids.\\

\begin{figure}
\resizebox{\hsize}{!}{\includegraphics[angle=-90]{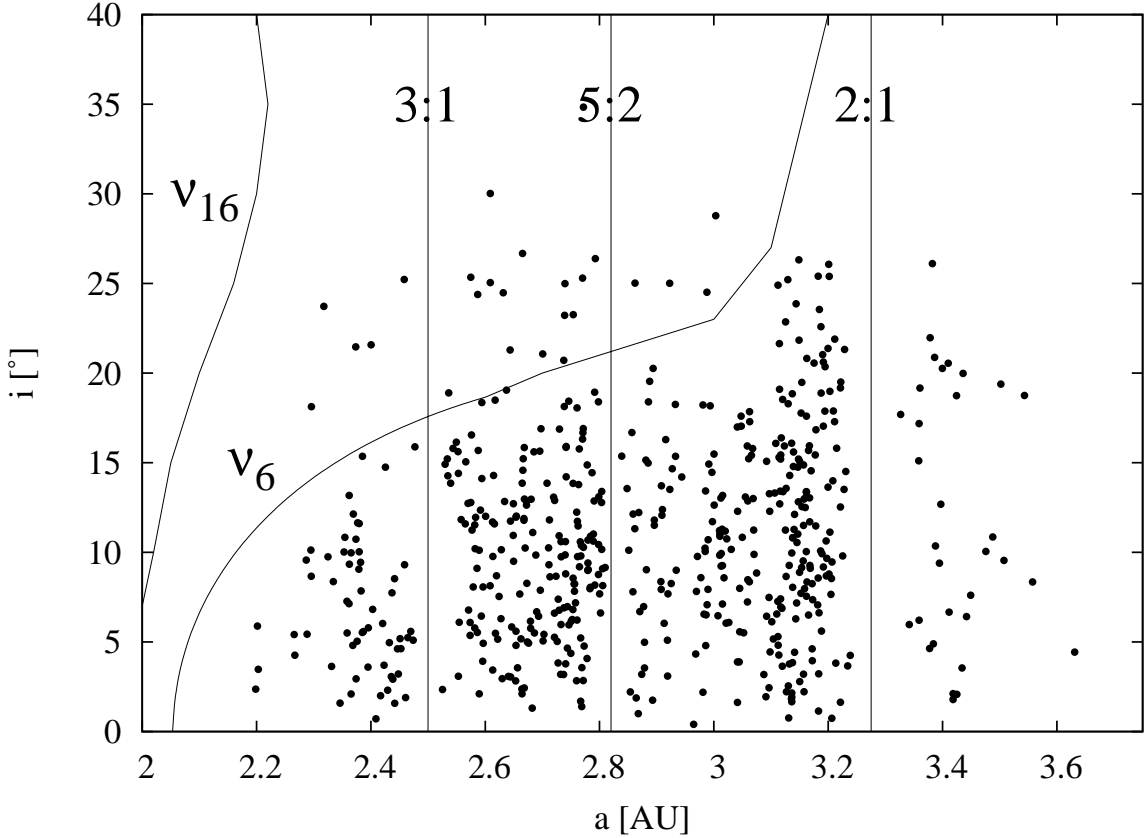}}
\caption{Current orbital distribution (inclination vs. semi major
  axis) of the asteroid belt. Only asteroids with absolute magnitude
  $H<9.7$ are plotted (corresponding to a diameter of approximately 50
  km). The locations of the $g=g_6$ (labelled $\nu_6$) and $s=s_g$
  ($\nu_{16}$) secular resonances and of some of the major mean-motion
  resonances with Jupiter (labelled $n:m$, standing for the ratio
  between the orbital period of Jupiter and of an asteroid) are also
  plotted.}
\label{abnow}
\end{figure}

In Fig.~\ref{abnow} we present the current orbital structure of the
asteroid belt with asteroids brighter than absolute magnitude $H=9.7$
i.e. with diameter larger than $D\sim50$~km. The orbital distribution
of these large asteroids is not affected by observational biases
(Jedicke et al., 2002) nor could it have been significantly modified
over aeons by non-gravitational forces or family formation
events. Therefore, these asteroids represent the belt's orbital
distribution since the time when the migration of the giant planets
ceased.\\

All the observed gaps correspond to the {\it current} locations of the
main mean-motion resonances with Jupiter and of the $\nu_6$
(i.e. $g=g_6$) secular resonance. These mean-motion resonances and the
$\nu_6$ secular resonance can remove asteroids from the belt because
they increase their eccentricities and the asteroids become planet
crossing. Therefore, it is not surprising that gaps in the asteroid
distribution are visible around the current locations of these
resonances. On the other hand, the $\nu_{16}$ (i.e. $s=s_6$) secular
resonance produces large changes in the asteroids' inclination.\\

As the giant planets migrated, the mean motion resonances with Jupiter
must have moved sun-ward towards their current positions. According to
most accepted models the radial migration of Jupiter is expected to
have covered 0.2-0.3~AU; thus the mean motion resonances should have
moved inwards by approximately 0.1~AU. The $\nu_6$ and $\nu_{16}$
secular resonances also moved sun-ward but their range of migration was
much larger than that of the mean motion resonances. In fact, if the
orbital separation of Jupiter and Saturn increased by more than
$\Delta_0 =1$~AU (again, as predicted by all models), the $\nu_6$
resonance swept the entire asteroid belt as it moved inwards from
4.5~AU to 2~AU. The $\nu_{16}$ resonance swept the belt inside of
2.8~AU (Gomes et al., 1997) to its current location at 1.9~AU. \\

We now determine the effect of planet migration on the orbital
distribution of the asteroid belt. We performed a sequence of
numerical simulations, using the swift-WHM integrator (Levison \&
Duncan, 1994; Wisdom \& Holman, 1991), as we explain here. We
proceeded in three steps. In the first step, the code was modified to
force the migration of Jupiter and Saturn as outlined in Paper~I:
the equations of motion were changed as to induce radial migration of
the planets, with a rate decaying as $\exp(-t/\tau)$.  For the reasons
explained in the previous section, the value of $\tau$ was set equal
to 5~My, the lower bound for $\tau_C$. \\

{Since the secular resonances only sweep the asteroid belt once the
period ratio between Saturn and Jupiter $P_S/P_J>2$ (Gomes, 1997) the
planets were started on orbits} with a period ratio of $P_S/P_J=2.03$:
Jupiter started at 5.40~AU and Saturn at 8.67~AU, similar to Malhotra
(1993) and Minton \& Malhotra (2009). The initial eccentricities and
inclinations of the giant planets with respect to the invariable plane
are close to their current values (Paper~I):
$(e_J,e_S)=(0.012,0.035)$ and $(i_J,i_S)=(0.23^\circ,1.19^\circ)$, and
so the strength of the secular resonances passing through the asteroid
belt does not change significantly throughout the migration. No
eccentricity damping was imposed on the giant planets. The terrestrial
planets were not included in this simulation, and neither were Uranus
and Neptune. This simulation covered a time-span of 25~My, i.e. 5 times
the value of $\tau$.\\

The primordial asteroid belt was situated between 1.8~AU and 4.5~AU
with orbits that did not cross those of Mars nor Jupiter. The initial
orbits of the asteroids were generated according to the following
recipe: take two random numbers {and assign them to the pericenter and
apocenter distances} on the interval $[1.8,4.5]$~AU. Then the
eccentricity and semi major axis of the asteroids are
$a\in(1.8,4.5)$~AU and $e\in(0,0.428)$. For the sake of simplicity the
inclination was chosen at random between 0 and 20$^\circ$ while the
three other angles were also chosen at random between 0 and
360$^\circ$. {Even though this method does not yield a uniform
distribution in semi major axis and eccentricity, it turns out that
this does not matter for the end result.} A total of 10\,000 asteroids
were used per simulation. We ran eight simulations altogether with
different choices of random numbers in the generation of the initial
conditions, for a total of 80\,000 test particles.  Asteroids were
removed if their distance to the Sun decreased below 1~AU (because in
reality they would be rapidly removed by encounters with the
Earth\footnote{We decided not to put a limit at 1.5~AU, corresponding
to Mars-crossing orbits, because the time scale for Mars encounters to
change significantly the asteroid's orbit is $\sim$100~My. Thus, in
principle, asteroids could become temporary Mars-crosser and then go
back into the main asteroid belt.}), when they entered the Hill sphere
of Jupiter, or they reached a distance further than 200~AU from the
Sun.\\

After this simulation, {we would like to continue the simulation
  including the effects of the terrestrial planets to account for the
  long-term modifications that these planets might have imposed to the
  asteroid belt orbital structure. Unfortunately this is not possible
  in a trivial way, because the final orbits of Jupiter and Saturn at
  the end of step 1 are not identical to the current orbits: the
  angular phases, for instance, are different. Thus, one cannot
  introduce other planets in the system, because the overall orbital
  evolution would then be different from the real dynamics of the
  solar system. Thus, we need to do first a simulation (step 2), where
  we bring} Jupiter and Saturn to their exact current orbits, while
  integrating the dynamical evolution of the asteroids that survived
  at the end of step 1. This was done by forcing the semi major axes,
  eccentricities and inclinations of Jupiter and Saturn to linearly
  evolve from the values at the end of step 1 to their current values,
  in 5~My; the same was done for the angular elements, assuming
  frequencies that were as close as possible to the current proper
  frequencies so that they performed the correct number of
  revolutions. The long time scale of 5~My allowed the asteroids to
  adapt in an adiabatic fashion to the new, slightly different
  configuration of the giant planets.  \\

At this point we could add the remaining planets to the system (the
terrestrial planets and Uranus and Neptune), assuming their current
orbital configuration. The third step of our study consisted in
following the asteroids for another 400~My under the influence of all
the planets. Now, asteroids were removed when they entered the Hill
sphere of any planet. \\

\begin{figure}
\resizebox{\hsize}{!}{\includegraphics[angle=-90]{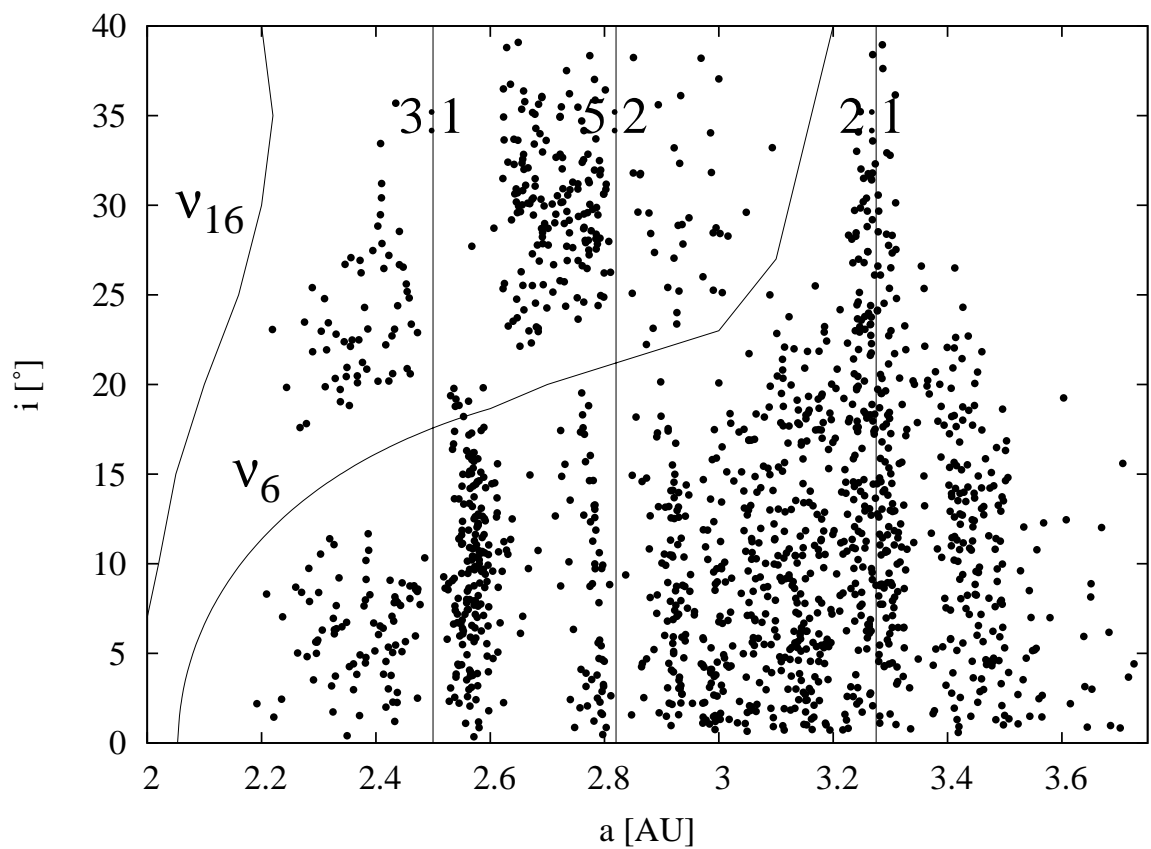}}
\caption{Orbital distribution of the asteroid belt at the end of the
  three step-wise simulations described in the main text.  The first
  of these simulations enacted planetesimal-driven migration of
  Jupiter and Saturn with a timescale $\tau=5$~My.}
\label{abaftermig}
\end{figure}

The final distribution of the asteroids at the end of these 3 steps is
depicted in Fig.~\ref{abaftermig}. It is clear that this distribution
is vastly different from Fig.~\ref{abnow}. The most striking
difference is that the ratio ${\cal R}$ between the number of
asteroids with inclinations above (denoted high-$i$) and below
(low-$i$) the $\nu_6$ resonance in the region of the asteroid belt
interior to 2.8~AU is much higher in Fig.~\ref{abaftermig} (0.7) than
in Fig.~\ref{abnow} (0.07). This ratio, at the beginning of the
simulation, was 0.08. \\

It is well-known that the effects of secular resonances on the orbital
distribution of the asteroids is anti-correlated with the speed at
which these resonances sweep through the asteroid belt (Nagasawa et
al., 2000). Thus, we believe ${\cal R}$ to be an important diagnostic
of the rate of planet migration. If the planets migrate slowly, as in
our simulation ($\tau=5$~My), the $\nu_{16}$ pushes asteroids to high
inclinations while the $\nu_6$ resonance is effective at removing the
low-inclination asteroids. This results in a high ${\cal R}$.  Planet
migrations with $\tau>5~My$ would give final distributions that are
likely even more different from the current belt than that presented
in Fig.~\ref{abaftermig}: the value of ${\cal R}$ would be either more
extreme or similar to what is presented here. In Minton \& Malhotra (2009)
the planets were assumed to migrate so fast ($\tau=0.5$~My) that 
the secular resonances swept the asteroid belt so
quickly that they did not have time to act; thus, the final value of
${\cal R}$ was similar to the initial one. \\

\begin{figure}
\resizebox{\hsize}{!}{\includegraphics[angle=-90]{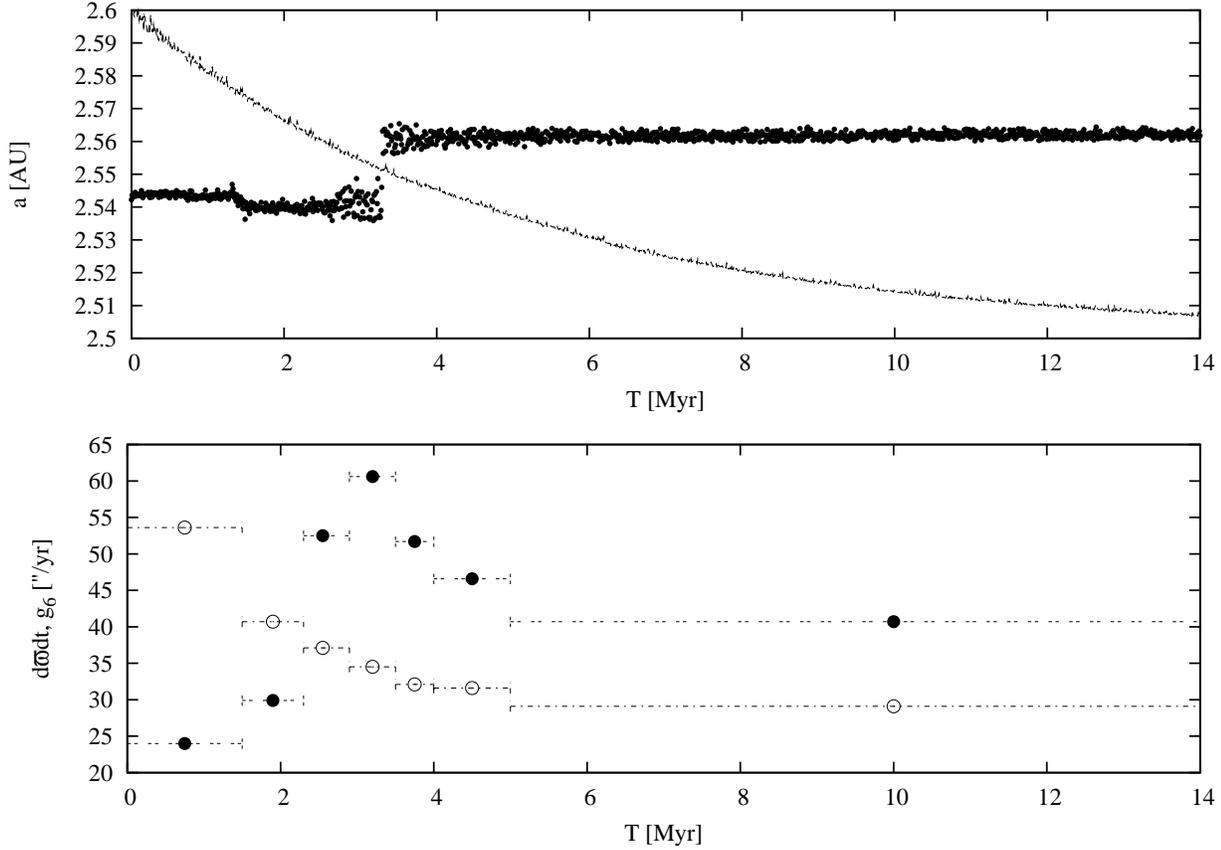}}
\caption{Example of an asteroid jumping across the 3:1 resonance with
Jupiter. The top panel depicts the semi major axis of the asteroid
(bullets) and the location of the 3:1 resonance with Jupiter (solid
line) as a function of time. The bottom panel shows the proper
precession frequency $\dot{\varpi}\equiv g$ of the asteroid (filled circles),
and $g_6$ (open circles) as a function of time. The values
are averages over the interval traced by the horizontal error bars.}
\label{3to1}
\end{figure}

The other two characteristic features of Fig.~\ref{abaftermig} that
are very different from the real asteroid distribution are (i) the
appearance of a dense group of objects at 2.55~AU, next to the current
location of the 3:1 resonance with Jupiter at 2.50~AU, and (ii) a
prominent lack of objects in the 2.6--2.7~AU range with inclinations
below the $\nu_6$ resonance.  The latter is caused by the sweeping of
$\nu_6$, which moves the asteroids to orbits with Earth-crossing
eccentricities, where they are removed. The clump next to the 3:1
resonance is formed by asteroids jumping across this resonance as the
latter moves sun-ward. An example of this event is depicted in
Fig.~\ref{3to1}. The top panel shows the semi major axis of an
asteroid (bullets) and the location of the 3:1 resonance with Jupiter
(solid line). One can clearly see the jump in semi major axis as the
asteroid encounters the resonance. Notice that the asteroid is not
captured in the resonance, but jumps from its inner side to the outer
side. The bottom panel of Fig.~\ref{3to1} shows (filled circles) the
average rate of the precession of the asteroid's longitude of
pericenter, $g$, over the time interval traced by the horizontal error
bars. As one can see, $g$ increases dramatically, by almost a factor
of 2, while the asteroid jumps across the resonance (Knezevic et al.,
1991). At the same time $g_6$ decreases smoothly and monotonically
(open circles). Thus, the asteroid crosses the $g=g_6$ resonance
extremely fast. This is not because $g_6$ decreases fast (as in Minton
and Malhtora, 2009), but because $g$ increases very fast. The result
is that, for asteroids jumping across the 3:1 resonance, the $g=g_6$
secular resonance sweeping is too fast to be effective, and therefore
these asteroids can survive on a moderate eccentricity orbit until the
end of the simulation. \\

All the results illustrated above strongly argue that the current
structure of the asteroid belt is incompatible with a smooth migration
of the giant planets with $\tau\sim 5$~My. As this is the minimal
time scale for planetesimal-driven migration, this excludes that the
orbital separation between Jupiter and Saturn increased due to the
sole process of planetesimal scattering. We think that there is no
easy way around this result for the following reasons:

\begin{itemize}
\item As explained above, larger values of $\tau$ beyond
 5~My, the only possible ones in planetesimal-driven migration,
 would not improve the final value of ${\cal R}$.\\

\item The original inclination distribution inside 2.8~AU is not
important for the final result because it is mixed by the sweeping of
the $\nu_{16}$ resonance; the inclinations are dispersed over the
interval from 0 to 30$^\circ$, whatever their initial distribution,
and thus the value of ${\cal R}$ is always nearly the
same. Figure~\ref{ab10} proves this claim. Here we show the result of a
smooth, planetesimal-driven migration simulation with $\tau=5$~My on
an asteroid belt with an initially uniform inclination distribution up
to 10$^\circ$ only. All other initial conditions are the same as in
the simulations discussed before. A visual comparison with the current
distribution in the asteroid belt (Fig.~\ref{abnow}) clearly shows
that there are still too many asteroids with semi major axes
$a<2.8$~AU above the $\nu_6$ resonance. In fact, for this simulation
${\cal R} = 0.6$, only slightly lower than in the original simulation of
Fig.~\ref{abaftermig} (0.7), but still an order of magnitude higher
than the observed ratio (0.07). {Thus, as we claimed above, inside
2.8~AU the initial inclination distribution does not matter for the
final result.}  Conversely, for $a> 2.8$~AU, the asteroid belt
preserves the initial inclination distribution because it is not swept
by the $\nu_{16}$ resonance. Consequently, we see in Fig.~\ref{ab10}
that there is a strong deficit of bodies with inclinations
$i>10^\circ$ beyond 2.8~AU, with the exception of the neighborhood of
the 2:1 mean-motion resonance with Jupiter (at 3.2~AU). As one can
see, this distribution does not match the observations either: a
broader initial inclination distribution would be required. \\

\begin{figure}
\resizebox{\hsize}{!}{\includegraphics[width=9cm]{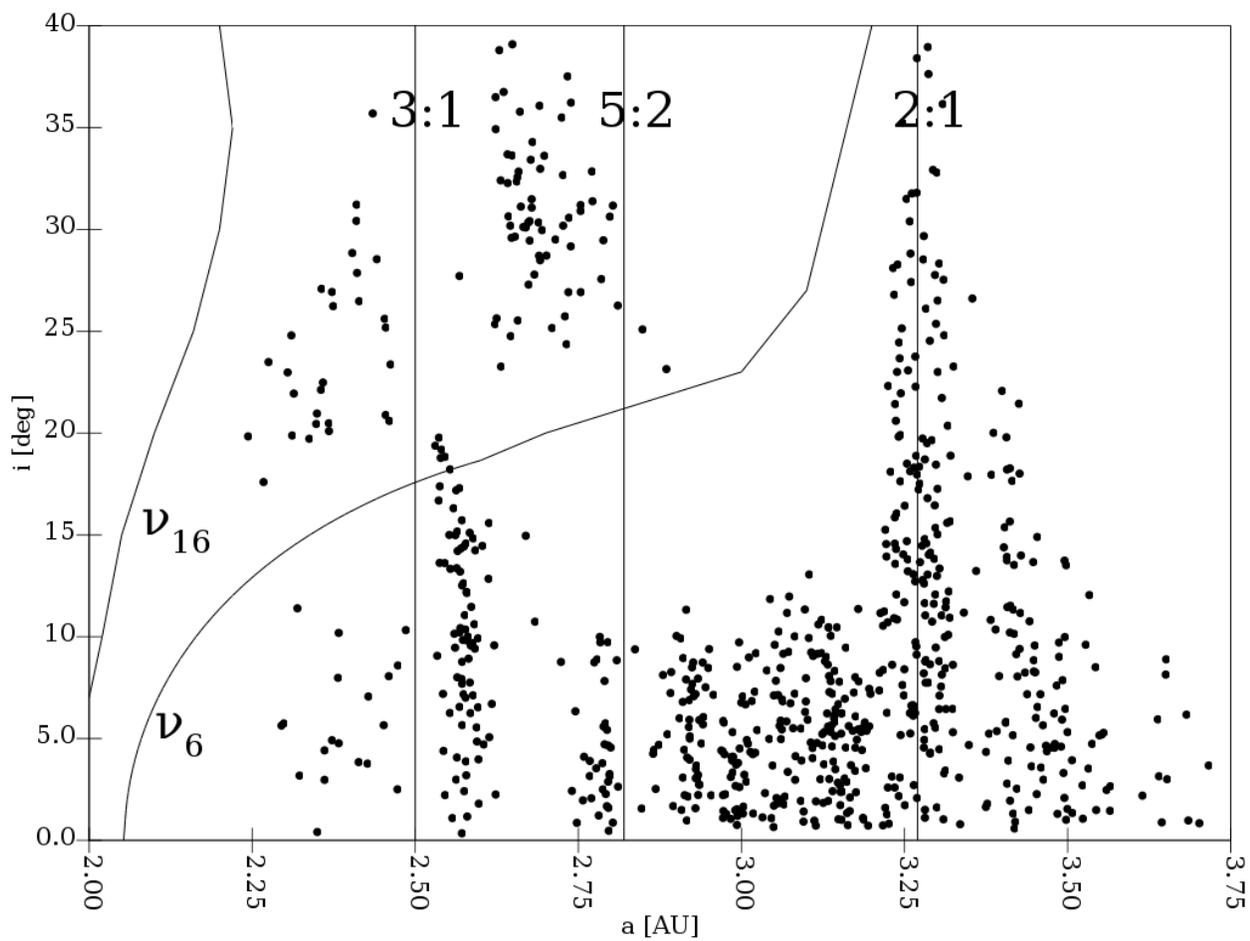}}
\caption{Orbital distribution of the asteroid belt after smooth
migration of Jupiter and Saturn with $\tau=5$~My, assuming the belt
had an original uniform inclination distribution up to 10$^\circ$.}
\label{ab10}
\end{figure}

\item All models agree that the separation between Jupiter and Saturn
increased by at least 1~AU, but it could have been more. Increasing
the distance travelled by Jupiter and Saturn would result in
practically the same dynamics because of two reasons: First, the
asteroid belt is only affected by secular resonances when $P_S/P_J>2$
(Gomes, 1997) i.e.\ towards the end of migration (Malhotra, 1993;
Tsiganis et al., 2005; Gomes et al., 2005). Hence, any prior migration
of the giant planets plays no role in shaping the belt.  Second, as we
said in sect.~\ref{migr}, the rate of migration of the resonances at a
given location (say in the inner asteroid belt) depends only on $\tau$
and is independent of the range of migration that the planets
travelled overall.\\

\item In principle one could envision that the eccentricities and/or
inclinations of Jupiter and Saturn were smaller than their current
values for most of the migration, thereby weakening the effects of the
secular resonances. However, this is unlikely because we are not aware
of any mechanism that is able to excite the eccentricities or
inclinations of the giant planets towards the end of the migration
(Paper~I). \\

\item The radial migration of the giant planets is believed to be the
last major event that sculpted the structure of the Solar System,
which coincided with the Late Heavy Bombardment (LHB) of the
terrestrial planets (Levison et al., 2001; Gomes et al., 2005; Strom
et al., 2005). Thus, we do not foresee any plausible mechanism capable
of subsequently erasing the signature of resonance sweeping in the
asteroid belt, particularly of decreasing ${\cal R}$ by an order
of magnitude. 
\end{itemize}

For all these reasons, we conclude that planetesimal-driven
migration could not have dominated the radial displacement of
Jupiter and Saturn. Thus below we look for an alternative mechanism
that results in a migration that is faster than the
planetesimal-driven one and hence is consistent with the short time
scale used by Minton \& Malhotra (2009).

\section{A jumping-Jupiter scenario}
\label{JJ}

Encounters between the giant planets, besides planetesimal scattering,
is the only mechanism that is known to us to be able to produce
large-scale orbital separation of the planets. However, planetary
encounters do not necessarily help our case of speeding up the
divergent migration between Jupiter and Saturn. For example, if Saturn
scatters an ice giant (Uranus or Neptune) outwards, it has to recoil
towards the Sun due to energy conservation. If Jupiter does not
encounter any planet, the orbital separation between Jupiter and
Saturn {\it decreases} and planetesimal-driven migration is still
required to bring the planets to their current orbital
separation. This would lead to the same problems with the asteroid
belt as discussed above. Thus, such a series of events has to be
rejected. However, if an ice giant is first scattered inwards by
Saturn and subsequently outwards by Jupiter, then the orbital
separation between Jupiter and Saturn increases abruptly. As in paper
II, we call this a `jumping-Jupiter evolution', in which the orbital
separation between Jupiter and Saturn increases on a time scale of
$10\,000$--$100\,000$ years, even shorter than assumed in Minton \&
Malhotra (2009). However, the motion does not follow a smooth,
exponential form. In the Nice model (Tsiganis et al., 2005; Gomes et
al., 2005) a `successful' jumping-Jupiter evolution, i.e. one where
Uranus and Neptune end on orbits with semi major axes within 20\% of
their current values, occurs in $\sim10\%$ of our simulations (Paper
II).\\

We stress that, at least in the Nice model, there is no appreciable
difference in the initial conditions of the jumping-Jupiter evolutions
with respect to those which lead to Jupiter not being involved in
encounters. This is because of the chaotic nature of the dynamics, due
to which both types of evolutions can originate from practically the
same simulation setup.\\

An example of a jumping-Jupiter evolution is shown in Fig.~\ref{jj}
and was taken from Paper II. The black and grey curves show the
evolution of the semi major axes of Jupiter and Saturn (top panel),
and their eccentricities (bottom panel), reported on the left-side and
right-side vertical scales, respectively. The stochastic behavior is
caused by repeated encounters with a Uranus/Neptune-mass planet (not
shown here) which was originally placed the third in order of
increasing distance from the Sun. Time $t=0$ is arbitrary and
corresponds to the onset of the phase of planetary instability. The
full evolution of the planets lasts 4.6 My and all giant planets
survived on stable orbits that are quite similar to those of the real
planets of the Solar System. The final ratio of the orbital periods of
Saturn and Jupiter is 2.45, very
close to the current value.  \\

\begin{figure}
\resizebox{\hsize}{!}{\includegraphics[angle=-90]{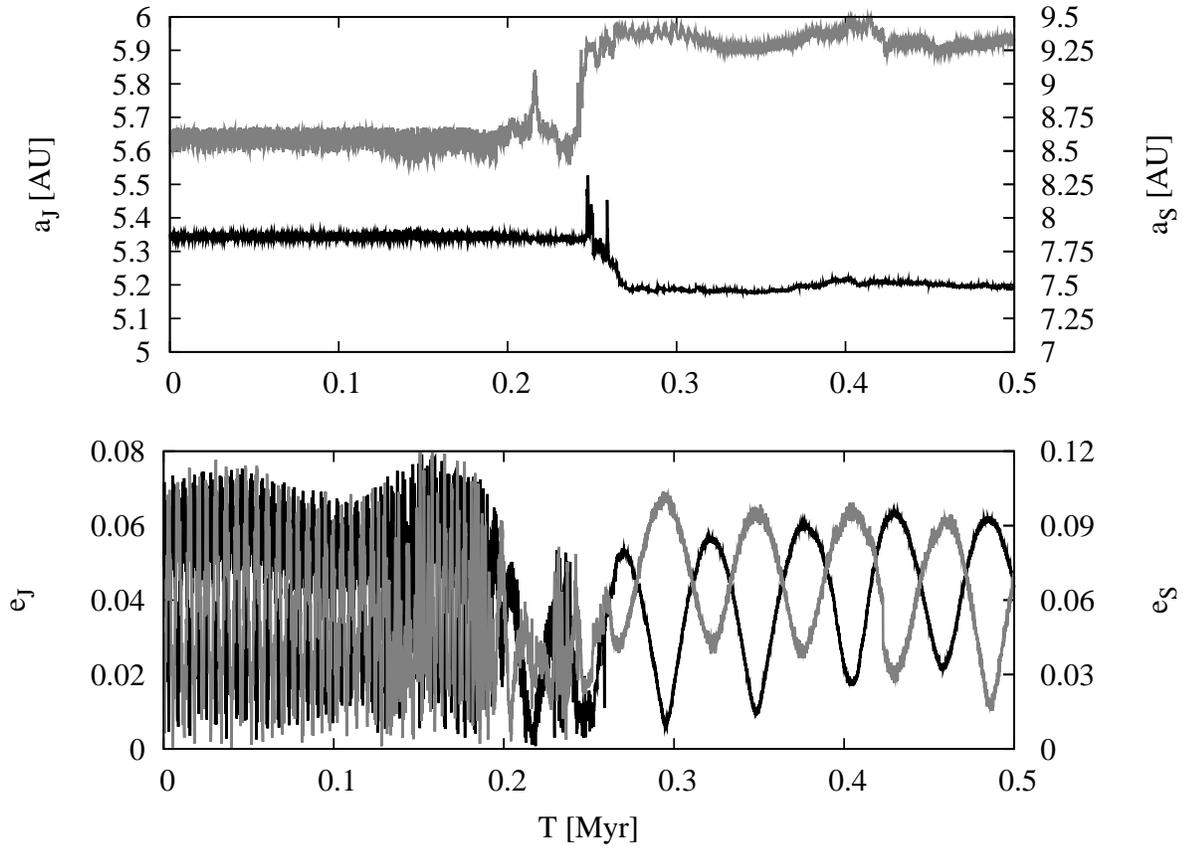}}
\caption{Example of a jumping-Jupiter evolution. The top panel depicts
the semi major axes of Jupiter (black) and Saturn (grey) -- reported
on the left and right sides respectively -- during the first 0.5~My
of the simulation. Their eccentricities are plotted in the bottom
panel.}
\label{jj}
\end{figure}

\section{Evolution of the asteroid belt during a jumping-Jupiter evolution}
\label{JJast}
In this section, we present results of the distribution of the
asteroid belt following the jumping-Jupiter scenario of
Fig.~\ref{jj}. Once again we employ three distinct steps, in the same
manner as described in sect.~\ref{AB-smooth} for the smooth
migration. The only difference is that for the first step we enact the
jumping-Jupiter evolution rather than adopting a smooth, exponential
migration. We used the same initial conditions for the asteroids as
those adopted for the smooth migration case of Fig.~\ref{abaftermig}
and enacted the jumping-Jupiter evolution using the modified version
of swift-WHM, presented and tested in Paper II. In this code, the
positions of Jupiter and Saturn have been computed by interpolation
from the 100~yr-resolved output of the evolution shown in
Fig.~\ref{jj}. We eliminated the asteroids that collided with the Sun
(perihelion distance smaller than the solar radius) or were scattered
beyond 200~AU. The terrestrial planets were not included. \\

\begin{figure}
\resizebox{\hsize}{!}{\includegraphics[angle=-90]{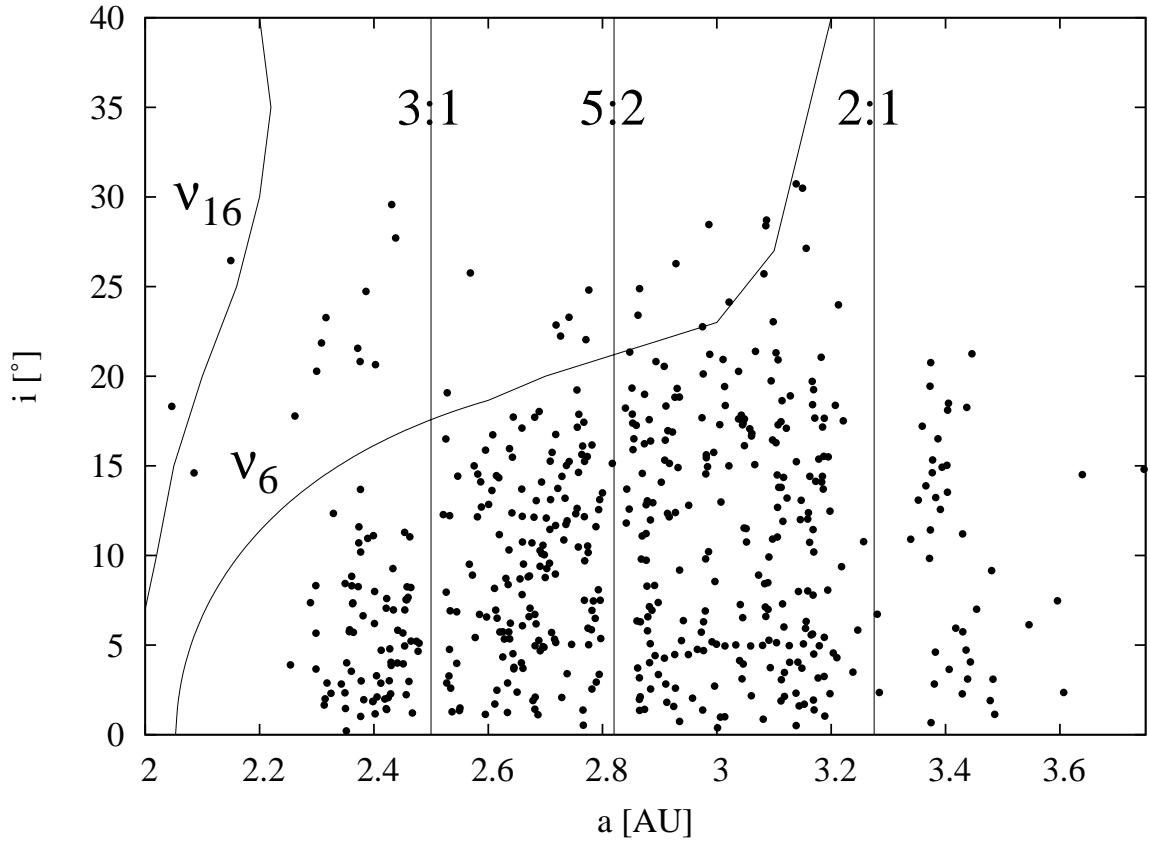}}
\caption{The orbital distribution of the asteroid belt 3.3 Gy after the end of the jumping-Jupiter evolution.}
\label{abjj}
\end{figure}

After the jumping-Jupiter evolution, we performed the second step in
the same manner as described in sect.~\ref{AB-smooth}, during which
Jupiter and Saturn are smoothly brought to their exact current
orbits. Finally, for the third step, the terrestrial planets and
Uranus and Neptune were added to the system on their current orbits
and with the correct phases, and the evolution of this system was
simulated for another 3.3~Gy. \\

The orbital distribution of the asteroid belt at the end of the third
step is shown in Fig.~\ref{abjj}. It is remarkably similar to the
current one depicted in Fig.~\ref{abnow}, with no spurious gaps and
clumps, unlike Fig.~\ref{abaftermig}. The final value of ${\cal R}$ is
identical to the observed value. However, unlike the smooth and slow migration
case, ${\cal R}$ now depends on the assumed initial inclination
distribution of the asteroids (here uniform up to 20$^\circ$). This is
because the $\nu_{16}$ resonance sweeps the belt so fast that it does
not significantly modify the inclinations of the asteroids, as
explained in sect.~\ref{AB-smooth}. \\

To illuminate this fact, in the framework of the same jumping-Jupiter
evolution, we simulated the evolution of asteroids that are subject to
the effects of sweeping by the $\nu_{16}$ resonance, i.e. asteroids
with $a < 2.8$~AU, but with inclinations up to 40$^\circ$. At the end
of this simulation, the root-mean-square change in inclination of the
asteroids was only 4$^\circ$, so that the original inclination
distribution is approximately preserved. Therefore, the
jumping-Jupiter scenario is consistent with the asteroid belt,
provided that the pre-migration asteroid distribution was similar to
the current inclination distribution of large asteroids, i.e.
relatively uniform in inclination, up to about 20$^\circ$. {An
orbital excitation of the asteroid belt pre-dating the jumping-Jupiter
evolution is likely. In fact, it is generally expected the asteroid
belt lost its original state of a dynamically cold disk during the
phase of terrestrial planets formation (Wetherill, 1992; Chambers and
Wetherill, 1998; Petit et al.,
2001; O'Brien et al., 2006, 2007). This happened much earlier than the
instability of the giant planets if the latter occurred at the LHB
time (see Appendix).} \\

As we said above, we are not aware of a third way that, after the
disappearance of the gas--disk, could widen the orbital separation
between Jupiter and Saturn, other than planetesimal-driven migration
and a jumping-Jupiter evolution. We have shown in
sect.~\ref{AB-smooth} that the former produces results inconsistent
with the structure of the asteroid belt, whatever its initial orbital
distribution. On the other hand, the jumping-Jupiter mechanism can work,
provided that the asteroid belt had the appropriate distribution, that
is an inclination distribution up to $\sim 20^\circ$, which is not
unreasonable. Thus, our conclusion is that the actual evolution of the
giant planets of the solar system was of the jumping-Jupiter
type. This conclusion has a long list of implications, which we
discuss in the next section.

\section{Implications}
\label{imply}

\subsection{Capture of the Irregular satellites of Jupiter}
Nesvorn\'{y} et al. (2007) showed that the capture of irregular
satellites is a generic outcome of planet-planet encounters and
reproduces their orbital distribution well. They showed that this
capture mechanism also works for Jupiter provided that it participates
in the encounters, which we support here.
Bottke et al. (2010) showed that the size distribution of the
irregular satellite systems is also consistent with this scenario.

\subsection{Contribution of main belt asteroids to the LHB}
To evaluate which fraction of the asteroid belt population becomes
unstable during the jumping-Jupiter evolution and what is the
subsequent overall asteroidal impact flux onto the Earth and the Moon,
we have proceeded with the same sequence of piece-wise simulations
similar to what was described in Sections~\ref{AB-smooth}
and~\ref{JJast}, but we added another step (step 0) prior to
re-enacting the jumping-Jupiter scenario (which is step 1).\\

For step 0 we started with the same uniform distribution of asteroids
as described in sect.~\ref{abaftermig}. We assumed that the giant
planets are situated on the orbits described in Morbidelli et
al. (2007). These orbits are fully resonant, quasi-circular and
co-planar.  Jupiter and Saturn are locked in their mutual 3:2 mean
motion resonance; Uranus is in the 3:2 exterior mean motion resonance
with Saturn and Neptune is in the exterior 4:3 mean motion resonance
with Uranus. These orbits are the result of hydro-dynamical
simulations of the dynamical evolution of the giant planets embedded
in a gas disk. This configuration is stable after the dispersal of the
gas nebula in absence of an exterior planetesimal disk, and are thus
believed to be the orbits the giant planets had in the interim between
the end of the gas disk phase and the onset of their global dynamical
instability. If one believes this instability triggered the LHB, this
compact configuration had to last for approximately 600~My.\\

To set up the right initial conditions for determining the amount of
mass transferred to the Moon, we have integrated the asteroids under
the influence of the giant planets on these presumed pre-LHB
multi-resonant orbits for 600~My (step 0). The terrestrial planets
were not included in this simulation. Asteroids were removed when
their perihelion distance decreased below 1.5~AU i.e. when they became
Mars-crossers, or had an encounter with Jupiter. A total of 724 out of
the original 1\,000 asteroids survived. The final orbital distribution in
semi major axis-eccentricity space of the asteroids is shown in the
{top-left panel} of Fig.~\ref{Renu}. As one
can clearly see, no gaps are visible near resonances with Jupiter. In
fact, it is well known that mean motion resonances are stable if
Jupiter follows a circular orbit (Morbidelli, 2002). This represents
what we believe is a realistic orbital distribution of the asteroids
at the onset of giant planet migration.\\

For step 1 (the second simulation), we integrated the remaining 724
particles while enacting the jumping-Jupiter evolution. The second
step was identical to what has been described earlier, while for the
third step the terrestrial planets, Uranus and Neptune were added and
the system was simulated for another 25~My. In the last three steps
asteroids were removed if they collided with the Sun, a planet or were
scattered beyond Jupiter. At the end of the third step 319 out of 724
particles survived (45\%). We expect that this number is close to the
final number of surviving asteroids because only a few were still
crossing the orbits of Mars and the Earth at the end of the last step.\\

To compute the mass impacting the Moon and the Earth we proceeded as
follows. During steps 1 and 2, we computed the collision
probability of every particle with our planet and its satellite,
assuming that the latter were on their current (fixed) orbits.
We performed this calculation using the algorithm
presented in Wetherhill (1967). During the last step, we used the same
approach, but adopted the orbital values of the Earth obtained during
the simulation itself. The Moon was assumed to be at the current
distance from our planet. \\

We understand that this procedure is not ideal. The fact that there
are no terrestrial planets in steps 1 and 2 probably increases the
lifetime of some objects that escaped from the asteroid belt. Also,
the strength of the $\nu_6$ resonance is severely reduced during step
2 {because the secular forcing between Jupiter and Saturn
is not taken into account}. This might also artificially enhance the
lifetime of some of the destabilized asteroids. For these reasons, our
results should be regarded as being upper bounds to the real flux of
matter from the asteroid belt towards the Earth-Moon system.\\

We find that the mean collision probability of our initial 724
asteroids with the Moon during steps 1 to 3 is $4\times 10^{-5}$. This
number includes particles on stable orbits, so the mean collision
probability of asteroids dislodged from the asteroid belt is about
twice as much i.e. $8 \times 10^{-5}$. The collision probability with
the Earth is 20 times larger. The mean impact velocity, before
gravitational focusing, is 20~km/s. \\

To translate these collision probabilities into estimates of the total
mass impacting the Earth-Moon system, we take into account that Minton
\& Malhotra (2010) argued that the asteroid belt might have lost half
of its population during the last 3.5~Gy because of chaotic
diffusion. The current mass of the asteroid belt is estimated to be
$3.6\times 10^{21}$~kg (Krasinsky et al., 2002) or $6 \times 10^{-4}$
Earth masses ($M_{\oplus}$). If the post-LHB mass was twice as much
and $\sim 50$\% of the population was destabilized during the
jumping-Jupiter evolution, then our 724 asteroids represent a total
mass of $\sim 1.5\times 10^{22}$~kg or $2.5 \times
10^{-3}$~$M_{\oplus}$. Multiplying this by the mean collision
probability with the Moon we find that a total of $6\times 10^{17}$~kg
was delivered to our satellite during the LHB from within the current
boundaries of the asteroid belt. The total mass delivered to the Earth
is 20 times larger. This is about an order of magnitude smaller than
the cometary contribution, as calculated in Gomes et al. (2005).  The
cometary flux is quite insensitive to the exact dynamics of the giant
planets, because the orbital evolution of the comets is dominated by
planetary scattering and not by resonance sweeping. {Thus,
unless the trans-Neptunian disk had a substantially lower mass than
currently thought, we have little reason} to believe that the
estimates about the cometary flux in Gomes et al. (2005) need to be
significantly revised in the framework of the jumping-Jupiter
scenario. Consequently, we conclude that, if the original asteroid
belt population was uniform in semi major axis, comets should have
dominated the LHB over main belt asteroids.\\

This conclusion, however, might violate constraints. There is a lively
debate on the nature of the LHB. It was argued (Kring \& Cohen, 2002)
that the basin-forming projectiles were neither comets nor primitive
asteroids, in contrast with our result. If this is confirmed, it will
be necessary to look for alternative sources of projectiles, such as
putative populations of small bodies in between the orbits of the
terrestrial planets, that could become fully destabilized during the
jumping-Jupiter evolution.

\subsection{The pre-LHB structure of the asteroid belt}
Since the jumping-Jupiter scenario preserves the original inclination
distribution, and only $\sim 50$\% of the main belt population is
removed onto planet-crossing orbits, the asteroid belt had to be
relatively similar to today's belt at the time when giant planet
migration started i.e. at the time of the LHB, if one accepts that the
two events are connected (Levison et al., 2001; Gomes et al., 2005;
Strom et al., 2005). More precisely, the asteroid inclinations had to
be spread between 0 and 20$^\circ$, and the total mass of the belt had
to be comparable to the current one. Thus, a very small number (if
any) of bodies larger than Ceres (called planetary embryos) could
exist. This provides a new set of formidable constraints for future
models of the primordial dynamical excitation and depletion of the
asteroid belt occurring during terrestrial planet formation
(Wetherill, 1992; Petit et al., 2001; O'Brien et al., 2006, 2007).

\subsection{Orbits of the giant planets before the onset of their dynamical instability}
The fact that the asteroid belt roughly preserves its pre-LHB
structure during a jumping-Jupiter evolution also provides constraints
on the orbits that the planets had to have had prior to their
migration. If the orbits of Jupiter and Saturn had had eccentricities
comparable to their current values, gaps would have been carved within
a few million years, exactly where the main resonances were located in
the pre-migration phase, and they would still be visible today. This
is demonstrated below. \\

Today, the eccentricity of Jupiter causes the main Jovian mean-motion
resonances to be unstable: the eccentricities of resonant asteroids
are increased to such large values that they begin to cross the orbits
of the planets (Gladman et al., 1997). Some of these planet-crossing
objects are removed from the resonances by encounters with these
planets, while others have their eccentricities increased to
Sun-grazing values. The removal of asteroids from these resonant
locations is the reason for the 'gaps' that we see in the distribution
of asteroids today (see Fig.~\ref{abnow}).\\

The most powerful resonance with Jupiter is the 3:1 resonance,
centered at 2.50~AU with a width of approximately 0.05~AU. If Jupiter
had had an eccentricity similar to its current value before migration,
when its semi major axis was about 5.4~AU, 
then a gap near 2.6~AU should have rapidly opened in the asteroid
distribution, clearly distinct from the currently observed gap at
2.5~AU. In fact, after the migration of Jupiter the gap opened by the
3:1 resonance at its final location would probably overlap with the
pre-migration gap, resulting in a wide empty zone ranging from
approximately 2.45 to 2.65~AU \\

\begin{figure}
\resizebox{\hsize}{!}{\includegraphics[angle=-90]{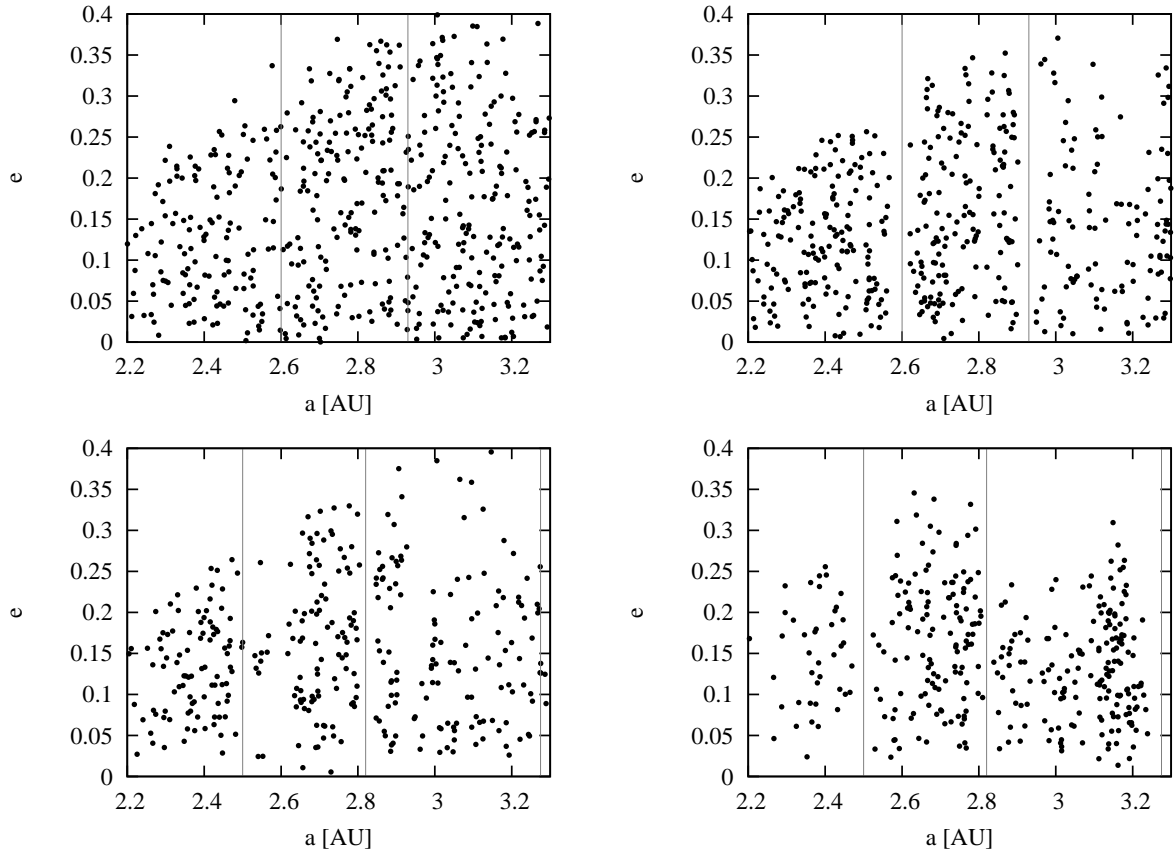}}
\caption{Semi major axis vs. eccentricity distribution of the asteroid
belt for various configurations of the giant planets. The top-left
illustrates the distribution after 600My of evolution with the giant
planets on multi-resonant quasi-circular orbits of Morbidelli et
al. (2007). The top-right panel shows the orbital
distribution after 600~My of evolution, assuming that Jupiter and
Saturn stayed on non-migrating orbits corresponding to the initial
conditions of Minton \& Malhotra (2009). The bottom-left panel
illustrates how the distribution in the top-right panel would have evolved
after a fast ($\tau=0.1$~My) exponential migration of Jupiter and
Saturn towards their current orbits, plus a 400~My subsequent
integration. For reference, the bottom right panel shows the current
distribution of 335 asteroids (same number as the particles plotted in
the bottom-left panel), among the real $H<9.7$ asteroid population.}
\label{Renu}
\end{figure}

We tested this hypothesis by performing the following
experiment. Jupiter and Saturn were placed on orbits similar to those
described in Malhotra (1993, 1995) and Minton \& Malhotra (2009):
Jupiter was set at 5.4~AU and Saturn at 8.7~AU on orbits with their
current eccentricities. There were no terrestrial planets. We
considered a distribution of asteroids consisting of 2\,000 test
particles, with the same initial conditions as described in
sect.~\ref{AB-smooth}. Assuming, as in Minton \& Malhotra (2009) and
Strom et al. (2005), that the trigger of giant planet migration was
delayed by $\sim 600$~My, so that it coincides with the beginning of
the LHB, we first simulated the evolution of these asteroids for
600~My, without imposing any migration on the giant planets. Once
again asteroids were removed when their perihelion distance decreased
below 1.5~AU or they encountered Jupiter. The asteroid distribution at
the end of this simulation is depicted in the {top-right}
panel of Fig.~\ref{Renu}. The gap associated with the 3:1 resonance
(located in this simulation at 2.6~AU) is clearly visible as well as
that of the 5:2 resonance (currently at 2.82~AU but here at
2.93~AU). These 'fossil' gaps are not observed today (c.f.\
Fig.~\ref{abnow}), so either they never formed or were refilled by
some mechanism (see below).\\

We subsequently simulated the remaining asteroids, imposing a smooth
exponential migration of the orbits of Jupiter and Saturn, but on a
time scale $\tau=0.1$~My. This choice of $\tau$ `mimics' the fast
migration time scale of the jumping-Jupiter scenario. After the
migration was finished, the system was simulated for another 400~My
without terrestrial planets, which is long enough for gaps to open
near the final positions of the resonances. The resulting orbital
distribution of the asteroid belt at this epoch is plotted in the
{bottom-left} panel of Fig.~\ref{Renu}. As expected, there
is a very large depletion of objects between 2.5 and 2.6~AU, which is
especially prominent at small eccentricities. A few objects remain in
the middle of this wide gap, because they managed to jump across the
sweeping 3:1 resonance. This wide gap is not observed in the asteroid
belt today. In fact, the {bottom-right} panel of Fig.~\ref{Renu}
shows, for comparison, the observed distribution of asteroids with
absolute magnitude $H<9.7$. For consistency, we have randomly selected
a number of real asteroids equal to the number of particles (335) that
remained until the end of the simulation, in the range of semi major
axes shown here. \\

We conclude from this experiment that migration on a jumping-Jupiter
time scale could not have replenished the fossil gaps opened by
Jupiter if the latter had been originally situated on an orbit with
eccentricity comparable to its current value.\\

We do not foresee any other mechanism that could replenish the fossil
gaps for the following reasons. First, remember that the real
asteroids that we are considering as tracers of the asteroid belt are
big (larger than approximately 50~km). These objects are too large to
migrate under the Yarkovsky effect. Additionally, only a few of them
are members of a collisional family. Thus, neither non-gravitational
forces, nor collisional breakups could have refilled the pre-migration
gaps with such large objects (the same is not true, obviously, for
smaller objects, for which we would not expect to see any fossil
traces of the pre-migration gaps).  Thus, the only remaining
possibility is that massive bodies (hereafter called planetary
embryos) resided in the belt and displaced the asteroids by
gravitational scattering. As we said above, it is unlikely that bodies
more massive than Ceres existed in the belt up to the time of giant
planet migration. But even if some embryo had been in the belt, it is
unlikely that it could have refill the gaps. In fact, from our first
simulation, where Jupiter and Saturn are at 5.4~AU and 8.7~AU with
their current eccentricities, we found that the mean lifetime of
particles with semi major axes in the range 2.57--2.61~AU (i.e.\ in
the region of the gap opened by the 3:1 resonance in the pre-migration
phase) is 6~My. This region is 25 times narrower (0.04~AU) than the
whole belt (1~AU). Thus, to prevent a gap from opening, embryos should
have been injecting 1/25 of all asteroids in this resonance every
6~My. Assuming that the pre-migration phase lasted 600~My, the
equivalent of 4 asteroid belts should have been injected into the 3:1
resonance, to be subsequently removed by its action. We strongly doubt
that any planetary embryo could have done this without leaving other
scars in the asteroid belt distribution, even more
pronounced than the fossil gaps themselves!\\

In conclusion, given that fossil gaps could not be replenished 
and they are not observed today, we believe that
it is unlikely that they ever existed. This implies that the giant
planets had to reside on quasi-circular orbits before the LHB, in
accordance with the hydro-dynamical simulations of Morbidelli et
al. (2007) {(c.f. the top-left panel of Fig.~\ref{Renu}).}\\

\section{Conclusions}
\label{concl}

Minton \& Malhotra (2009) showed that the current orbital structure of
the asteroid belt is consistent with a primordial uniform distribution
of asteroids in semi major axis, eccentricity and inclination space,
and an exponential migration of the giant planets with a
characteristic time scale $\tau=0.5$~My. However, they did not
demonstrate that $\tau$ needs to be this short, nor they discussed 
by which mechanism planet migration could operate on this timescale. \\

In this paper we have demonstrated that {
the migration of the giant planets, with $\tau=5$~My or longer,} is
inconsistent with the current structure of the belt, whatever its
initial orbital distribution. The divergent migration of Jupiter and
Saturn needs to be as fast as used in Minton \& Malhotra, or even
faster. \\

This constraint on the time scale of migration unveils the dominant
process by which the orbits of Jupiter and Saturn separated from each
other. In principle, two mechanisms are possible: (i) planetesimal
driven migration or (ii) encounters of both Jupiter and Saturn with a
third planet (presumably Uranus or Neptune). The first mechanism
results in an exponential migration, but the characteristic $\tau$
cannot be faster than 5~My. Given the incompatibility of this value
of $\tau$ with the current structure of the asteroid belt, this
process can be excluded. The second mechanism, dubbed 'jumping-Jupiter
evolution', results in a radial displacement of the orbits of Jupiter
and Saturn on a time scale shorter than 0.1~My, but the migration of
the planets cannot be well represented by an exponential law. We have
simulated the evolution of the asteroid belt during a jumping-Jupiter
evolution and we have found that the final orbital structure of the
belt matches well the observed one, provided that the original belt
had an approximately uniform inclination distribution that extended up
to $\sim 20^\circ$. \\

This result strongly argues that the real evolution of the giant
planets was of the jumping-Jupiter type and that the orbits of the
giant planets, before migration, were significantly more circular than
the current ones.\\

\section{Appendix: A temptative timeline of events characterising the Solar
  System evolution, as emerging from this and previous works.}

{The first solids of the solar system condensed 4.568 Gy ago (see
  Kleine et al., 2009 for a review). This is generally considered as
  {\it time zero} in solar system history.

  Because most of the mass of
  Jupiter and Saturn is in H and He, these planets must have formed in
  a quite massive gaseous proto-planetary disk (Pollack et al.,
  1996). Because proto-planetary disks typically last 3-5 My around
  young stars (Haisch et al., 2001), Jupiter and Saturn should have
  formed within this time. 

  When embedded in a gas disk, giant planets tend to migrate towards
  the central star (Lin and Papaloizou, 1986). Hydro-dynamical
  simulations (Masset and Snellgrove, 2001; Morbidelli and Crida,
  2007; Pierens and Nelson, 2008) have consistently shown that Saturn
  migrates faster than Jupiter and, as it approaches the latter, it is
  eventually captured in their mutual 2:3 mean motion resonance. At
  this point, the inward migration of the two planets stops. Depending
  on the disk parameters, the two planets either stay in place, or
  migrate outward in resonance, until the disappearance of the disk
  (Morbidelli and Crida, 2007). This explains why we do not have a hot
  Jupiter in our Solar System. Walsh et al. (2010) proposed that
  Jupiter migrated down to 1.5 AU before Saturn formed and was
  captured in resonance; when this occurred, the two planets reversed
  migration and Jupiter reached 5.4 AU when the gas-disk
  disappeared. Either way (i.e with or without a substantial outward
  migration of Jupiter and Saturn), when the gas disk disappeared
  around 3--5 My, Jupiter and Saturn should have been in their mutual
  2:3 mean motion resonance, with quasi circular and co-planar orbits,
  with Jupiter at about 5.4~AU. Uranus and Neptune, should also have
  been in resonances with Saturn and with themselves, trapped in this
  configuration by differential migration during the gas-disk phase:
  however, there are multiple resonances in which these planets might
  have been (Morbidelli et al., 2007; Batygin and Brown, 2010).

At the disappearance of the gas, two massive disks of planetesimals
should have remained: one in the inner solar system, with an outer
edge located close to the minimal orbital radius that Jupiter acquired
during its evolution in the gas-disk, and one in the outer solar
system, beyond the orbit of the most distant planet. The terrestrial
planets formed by mutual collisions from the inner disk. Geochemical
constraints (Touboul et al., 2007; Allegre et al., 2008; Kleine et
al., 2009) and numerical modeling (Chambers and Wetherill, 1998;
Chambers, 2001; Raymond et al., 2006; O'Brien et al., 2006; Hansen,
2009; Morishima et al., 2010) suggest that the terrestrial planets
took 30--100~My to form.

By the time terrestrial planet formation was completed, the asteroid
belt was substantially depleted and dynamically excited.  If Jupiter
never went significantly inwards of 4--5 AU, the depletion and
dynamical excitation of the asteroid belt occurred during the
terrestrial planets formation process, due to the combination of
perturbations from resident planetary embryos and Jupiter (Wetherill,
1992; Chambers and Wetherill, 1998; Petit et al., 2001; Raymond et
al., 2006; O'Brien et al., 2006, 2007). In the Walsh et al. (2010)
scenario, the asteroid population had already been depleted and
excited by the inward and outward migration of Jupiter through the
main belt region during the gas-disk phase. The orbital excitation of
the asteroid belt at $t\sim 100$~My is model dependent. For instance,
in Petit et al. (2001) the final inclinations of the surviving
asteroids were within 15 degrees; in O'Brien et al. (2006) most of the
surviving asteroids had $10^\circ<i<30^\circ$; in Walsh et al. (2010)
the resulting inclination distribution was roughly uniform up to 20
degrees, in agreement with the findings of the present paper.

In the mean time, the outer planetesimal disk was slowly grinding into
dust by collisional comminution, losing about a factor of 2 in mass in
600~My (Booth et al., 2008). The gravitational interactions between
the giant planets and this disk slowly modified the orbits of the
former, eventually extracting the planets from their mutual mean
motion resonances (Morbidelli et al., 2007; Batygin and Brown, 2010;
Levison et al., in preparation). At this point the giant planets
became violently unstable.  The occurrence of a Late Heavy Bombardment
on the terrestrial planets $\sim 3.9$~Gy ago, strongly suggest that
this transition to instability occurred $\sim 650$~My after {\it time
zero} (Levison et al., 2001; Gomes et al., 2005; Strom et al., 2005),
i.e. well after the formation of the terrestrial planets and the
depletion/excitation of the asteroid belt. Due to this instability,
the giant planets started to have mutual encounters; Uranus and
Neptune were scattered into the outer planetesimal disk and dispersed
it. As a result of mutual scattering among the planets and dynamical
friction exerted by the outer disk, the planets finally acquired their
current orbits (Thommes et al., 1999, 2000; Tsiganis et al., 205;
Morbidelli et al., 2007; Batygin and Brown, 2010; Levison et al., in
preparation).

The present work strongly argues that, of all possible dynamical paths
that the planets could have followed from their original
multi-resonant orbits to the current configuration, the real one had
to be of the 'jumping-Jupiter' type. In other words the orbital separation
between Jupiter and Saturn had to increase abruptly as a result of
encounters of both these planets with either Uranus or
Neptune. Consequently, the orbits of the terrestrial planets (Brasser
et al., 2009) and of the asteroid belt (this work) were only
moderately affected by the giant planet instability.

The objects from the outer planetesimal disk and asteroids
escaping from the main belt both contributed to the Late Heavy
Bombardment of the terrestrial planets, with the former dominating
over the latter (this work). However, it is possible that other
asteroid belts existed in regions that were stable before the giant
planet instability and became unstable since; objects from these belts
might have dominated the LHB, but this possibility needs further
investigation to be supported. A small fraction of the planetesimals
of the outer disk survived on stable trans-Neptunian orbits,
corresponding to today's Kuiper belt (Levison et al., 2008).

Thus, the Solar System acquired its current global structure at the
LHB time and did not evolve substantially since then.}

}

\noindent{\bf Acknowledgements:}

This work is part of the Helmholtz Alliance's 'Planetary evolution and
Life', which RB and AM thank for financial support. Exchanges between
Nice and Thessaloniki have been funded by a PICS programme of France's
CNRS. HFL is grateful to NASA's Origins of Solar Systems and Outer
   Planets Program for funding.

\section{References}

All{\`e}gre, C.~J., Manh{\`e}s, G., Gopel, C.\ 2008, Earth and Planetary Science Letters, 267, 386 \\

Bailey, B.~L., \& Malhotra, R.\ 2009, Icarus, 203, 155 \\

Batygin, K., \& Brown, M.~E.\ 2010, \apj, 716, 1323 \\

Booth, M., Wyatt, M.~C., 
Morbidelli, A., Moro-Mart{\'{\i}}n, A., Levison, H.~F.\ 2009.\  Monthly Notices of the Royal Astronomical Society 399, 385-398. \\

Bottke, W.~F., 
Nesvorn{\'y}, D., Vokrouhlick{\'y}, D., 
\& Morbidelli, A.\ 2010, \aj, 139, 994 \\

Brasser, R., Morbidelli, A., Gomes, R., Tsiganis, K., \& Levison, H.~F.\ 2009, \aap, 507, 1053 \\

Chambers, J.~E., \& Wetherill, G.~W.\ 1998, Icarus, 136, 304 \\

Chambers, J.~E.\ 2001, 
Icarus, 152, 205 \\

Di Sisto, R.~P., \& Brunini, A.\ 2007, Icarus, 190, 224 \\

Fernandez, J.~A., \& Ip, W.-H.\ 1984, Icarus, 58, 109 \\

Gladman, B.~J., et al.\ 
1997, Science, 277, 197 \\

Gomes, R.~S.\ 1997, \aj, 114, 
396 \\

Gomes, R.~S., Morbidelli, 
A., \& Levison, H.~F.\ 2004, Icarus, 170, 492 \\

Gomes, R., Levison, 
H.~F., Tsiganis, K., \& Morbidelli, A.\ 2005, \nat, 435, 466 \\

Hahn, J.~M., \& Malhotra, R.\ 1999, \aj, 117, 3041 \\

Haisch, K.~E., Jr., 
Lada, E.~A., Lada, C.~J.\ 2001.\ The Astrophysical Journal 553, L153-L156. 

Hansen, B.~M.~S.\ 2009.\ The 
Astrophysical Journal 703, 1131-1140. \\

Ida, S., Bryden, G., Lin, 
D.~N.~C., \& Tanaka, H.\ 2000, \apj, 534, 428 \\

Jedicke, R., Larsen, 
J., \& Spahr, T.\ 2002, Asteroids III, 71 \\

Kirsh, D.~R., Duncan, M., 
Brasser, R., \& Levison, H.~F.\ 2009, Icarus, 199, 197 \\

Kleine, T., Touboul, M., 
Bourdon, B., Nimmo, F., Mezger, K., Palme, H., Jacobsen, S.~B., Yin, Q.-Z., 
Halliday, A.~N.\ 2009.\  Geochimica et Cosmochimica 
Acta 73, 5150-5188. \\

Knezevic, Z., Milani, 
A., Farinella, P., Froeschle, C., \& Froeschle, C.\ 1991, Icarus, 93, 316 \\

Krasinsky, G.~A., 
Pitjeva, E.~V., Vasilyev, M.~V., \& Yagudina, E.~I.\ 2002, Icarus, 158, 98 \\

Kring, D.~A., \& Cohen, B.~A.\ 2002, Journal of Geophysical Research (Planets), 107, 5009 \\

Levison, H.~F., \& Duncan, M.~J.\ 1994, Icarus, 108, 18 \\

Levison, H.~F., Dones, 
L., Chapman, C.~R., Stern, S.~A., Duncan, M.~J., 
\& Zahnle, K.\ 2001, Icarus, 151, 286 \\

Levison, H.~F., 
Morbidelli, A., Vanlaerhoven, C., Gomes, R., Tsiganis, K.\ 2008.\ Icarus 196, 258-273. \\

Lin, D.~N.~C., 
Papaloizou, J.\ 1986.\  The 
Astrophysical Journal 309, 846-857. \\

Malhotra, R.\ 1993, \nat, 
365, 819 \\

Malhotra, R.\ 1995, \aj, 110, 
420 \\

Masset, F., 
Snellgrove, M.\ 2001.\ Monthly Notices of the Royal Astronomical 
Society 320, L55-L59. \\

Minton, D.~A., \& Malhotra, R.\ 2009, \nat, 457, 1109 \\

Minton, D.~A., \& Malhotra, R.\ 2010, Icarus, 207, 744 \\

Morbidelli, A. 2002 Modern Celestial Mechanics - Aspects of Solar System Dynamics (Taylor \& Francis, UK).\\

Morbidelli, A., 
Crida, A.\ 2007.\ Icarus 191, 158-171. \\

Morbidelli, A., 
Tsiganis, K., Crida, A., Levison, H.~F., \& Gomes, R.\ 2007, \aj, 134,
1790 \\

Morbidelli, A., Brasser, R., Tsiganis, K., Gomes, R., \& Levison, H.~F.\ 2009, \aap, 507, 1041 \\

 Morishima, R., 
Stadel, J., Moore, B.\ 2010.\ 
Icarus 207, 517-535. \\

Nagasawa, M., Tanaka, 
H., \& Ida, S.\ 2000, \aj, 119, 1480 \\

 Nesvorn{\'y}, D., 
Vokrouhlick{\'y}, D., \& Morbidelli, A.\ 2007, \aj, 133, 1962 \\

O'Brien, D.~P., 
Morbidelli, A., \& Levison, H.~F.\ 2006, Icarus, 184, 39 \\

O'Brien, D.~P., 
Morbidelli, A., \& Bottke, W.~F.\ 2007, Icarus, 191, 434 \\

 Petit, J.-M., Morbidelli, 
A., \& Chambers, J.\ 2001, Icarus, 153, 338 \\

 Pierens, A., Nelson, R.~P.\ 2008.\  Astronomy and Astrophysics 482, 333-340.\\

 Pollack, J.~B., 
Hubickyj, O., Bodenheimer, P., Lissauer, J.~J., Podolak, M., Greenzweig, 
Y.\ 1996.\  Icarus 124, 62-85.\\

 Raymond, S.~N., Quinn, 
T., Lunine, J.~I.\ 2006.\ Icarus 183, 
265-282. \\

Strom, R.~G., Malhotra, 
R., Ito, T., Yoshida, F., \& Kring, D.~A.\ 2005, Science, 309, 1847 \\

Thommes, E.~W., Duncan, 
M.~J., \& Levison, H.~F.\ 1999, \nat, 402, 635 \\

Thommes, E.~W., Duncan, 
M.~J., \& Levison, H.~F.\ 2002, \aj, 123, 2862 \\

Touboul, M., Kleine, 
T., Bourdon, B., Palme, H., \& Wieler, R.\ 2007, \nat, 450, 1206 \\

Tsiganis, K., Gomes, 
R., Morbidelli, A., \& Levison, H.~F.\ 2005, \nat, 435, 459 \\

 Walsh, K., Morbidelli, A., Raymond, S., O'Brien, D. and Avi,
  M. 2010. DPS abstract, 2010.\\

Wetherill, G.~W.\ 1967, 
\jgr, 72, 2429 \\

Wetherill, G.~W.\ 1992, 
Icarus, 100, 307 \\

Wisdom, J., \& Holman, M.\ 1991, \aj, 102, 1528  \\

\end{document}